\documentstyle[12pt,openbib]{article}
\title{{\bf Universal scaling in BCS superconductivity in three dimensions in 
non-$s$ waves}\thanks{To Appear in European Physical Journal B}}

\author{Angsula Ghosh and Sadhan K. Adhikari \\
Instituto de F\'{\i}sica Te\'orica, Universidade Estadual Paulista\\
01.405-900 S\~ao Paulo, S\~ao Paulo, Brazil\\}

\date{\today}

\begin{document}
\maketitle

\begin{abstract}

The solutions of a renormalized BCS equation are studied in three space
dimensions in $s$, $p$ and $d$ waves for finite-range  separable potentials
in the weak to medium coupling region. In the weak-coupling limit, the
present  BCS model yields a small  coherence length $\xi$ and a
large critical temperature, $T_c$, appropriate for some high-$T_c$ materials.
The BCS gap, $T_c$, $\xi$ and  specific heat  $C_s(T_c)$ as a function of
zero-temperature condensation energy are found to exhibit 
potential-independent universal scalings.
   The entropy, specific heat,
spin susceptibility and penetration depth as a function of temperature
exhibit universal scaling below $T_c$ in $p$ and $d$
waves.

PACS Numbers: {74.20.Fg, 74.72.-h}


\end{abstract}

\section{Introduction}

At  low temperature, a collection of weakly interacting electrons
spontaneously form  large overlapping Cooper pairs \cite{e} according to the
microscopic Bardeen-Cooper-Schreiffer (BCS) theory of superconductivity
\cite{8,t}. There has been renewed interest in this problem with the
discovery of enhanced  superconductivity in alkali-metal-fulleride compounds
\cite{f} and cuprates \cite{dw}. The fulleride compounds, with critical
temperature $T_c$ up to $\sim 30-40$K, exhibit superconductivity in three
dimensions and have a relatively small coherence length $\xi: \xi k_F \sim 10
- 100$, with $k_F$ the Fermi momentum.  
At zero temperature $\xi$ is essentially the pair radius.
In the application of the BCS theory to high-$T_c$ materials, the
serious challange  is to consistently produce a large $T_c$ and a small $\xi
$ in the weak-coupling region. The usual  phonon-induced BCS model is unable
to produce  a large $T_c$  in the weak-coupling region.

Despite much effort, the normal state of  the high-$T_c$ superconductors
has not been satisfactorily understood.  Unlike the conventional
superconductors, their normal state exhibits peculiar
properties. The thermodynamic and electromagnetic observables of these
materials above $T_c$ have temperature dependencies which are very different
from those of a Fermi liquid
\cite{n}. 
There are  controversies about the appropriate microscopic hamiltonian,
pairing mechanism, and gap parameter for them\cite{n,c}.

The  BCS theory considers $N$ electrons of spacing $L$, interacting via a
weak potential of short range $r_0 $ such that $r_0 << L$ and $r_0 << \xi$.
When suitably scaled, most  properties of the system should be insensitive to
the details of the potential and be universal functions of the dimensionless
variable $L/\xi$ \cite{L}.  
In this work we
study the weak-coupling BCS problem in three dimensions for $s$, $p$, and $d$
waves with two objectives in mind. The first is to identify the
universal nature of the solution appropriate to high-$T_c$ superconductors.
We would  specially be interested to find out if the weak-coupling 
BCS theory can
explain some of the universal behaviors of high-$T_c$ materials independent
of the above-mentioned controversies.
The second objective is to find out to
what extent the universal nature of the solution is modified in the presence
of realistic finite-range (nonlocal separable) potentials.  Instead of
solving the BCS equation on the lattice with appropriate symmetry, we solved
the equations in the continuum. This procedure should suffice for  present
objectives.   There is also the
possibility of Cooper pairing in non-$s$ waves, such as, $p$-wave pairing in
superfluid $^3$He \cite{b,a} and $d$-wave pairing in some superconductors
\cite{dw}. Hence 
the present discussion of universality  is also extended to $p$ and $d$
waves.

In place of the standard phonon-induced BCS model we employ a renormalized
BCS model in three dimensions with separable and zero-range potentials,
 which has
certain advantages. 
The  standard BCS model yields the following linear
correlation between $T_c$ and $T_D$, where $T_D$ is the Debye temperature:
$T_c \approx 1.13 T_D \exp (-1/\bar \lambda)$ \cite{t}, where $\bar
\lambda$ is
the effective strength of the phonon-induced BCS interaction. Due
to the above correlation with $T_D$, $T_c$ of the standard BCS model is low.
This correlation between $T_c$ and $T_D$ is fundamental in explaining the
observed isotope effect in conventional superconductors \cite{t}. The
high-$T_c$ materials exhibit a very reduced and negligible isotope effect and
a visible linear correlation between $T_c$ and $T_F$ \cite{u}, where $T_F$ is
the Fermi temperature.    The present renormalized BCS
model yields a linear scaling between $T_c$ and $T_F$. Because of this
scaling with $T_F$,  the present $T_c$
can be large and appropriate for the high-$T_c$ materials in the weak-coupling
region.  In addition, the present model also produces an appropriate
$T_c/T_F$ ratio and a small $\xi$ in the weak-coupling region in accord with
recent experiments \cite{u} on high-$T_c$ materials.

Previously, there have been studies of the solution of BCS equations in
terms of potential strength, $V_0$,  or the pair 
scattering length in vacuum, $a$,
employing a  short-range potential \cite{No}.  Such studies
have not revealed the universal nature of the transition from weak-
to medium-coupling. Here, we employ the
zero-temperature condensation energy per particle, $\Delta U$, of the BCS
condensate  as the reference variable for studying the problem. As
$\Delta U$  increases, one passes from weak to medium coupling. We
calculate  the zero-temperature BCS gap  $\Delta(0)$, $T_c$, the specific
heat per particle $C_s(T_c)$  in different 
 partial waves and the zero-temperature
pair size $\xi$ in $s$ wave. These observables obey robust
universal scaling as functions of $\Delta U$  valid over several decades in
the weak-coupling region independent of the  range of potential.
Similar scalings were not found when  $\Delta(0)$, $T_c$, $C_s(T_c)$ and
$\xi$ were considered as a function of $V_0$ or $a$ as in Ref.
\cite{No}.

We also  calculate the temperature dependencies of different quantities, such
as, the BCS gap  $\Delta(T)$,  penetration depth $\lambda_s(T)$,
spin-susceptibility $\chi_s(T)$, $C_s(T)$, internal energy per
particle $U_s(T)$ and
entropy $S_s(T)$ for $T<T_c$.  Of these, the $T$ dependencies of $S_s(T)$, 
$C_s(T)$,
$\chi_s(T)$, and $\lambda_s(T)$ are interesting.  For
isotropic $s$ wave,  the BCS theory yields exponential dependence on
temperature as $T \to 0$ for  these observables independent of space
dimension \cite{t,t2}.  The observed power-law dependence on temperature in
some of these quantities  \cite{t2,s,t1}
can  be explained with anisotropic gap function in
non-$s$ waves with node(s) on the Fermi surface.  We find  universal power-law
dependence in non-$s$ waves independent of the range or strength of
potential.  For $l\ne 0$ we find
\begin{eqnarray}
S_s(T)/  S_s(T_c) &\approx &(T/T_c)
^{\beta_S}\label{a}\\
C_s(T)/ C_n(T_c)  &\approx & D(T/T_c)^{\beta_C}\\
\chi_s(T)/\chi_s(T_c) &\approx & (T/T_c)^{\beta_\chi}\\ 
\Delta \lambda (T)\equiv(\lambda_s(T)-\lambda_s(0))/\lambda_s(0)
 &\sim & (T/T_c)^{\beta_\lambda},\label{d}
\end{eqnarray}
  valid for a wide range of temperature.
  The suffix $n$ and $s$  refer to normal and
superconducting states, respectively. Similar power-law dependencies were
predicted from an analysis of experimental data \cite{s} as well as from a
calculation based on Eliashberg equation \cite{t1}.

From the weak-coupling BCS theory we established the
following  relations analytically: 
$\Delta(0)/\sqrt{\Delta U} = \sqrt{8/3},$
$T_c/\sqrt{\Delta U}\-= \-\sqrt{8/3} A^{-1},$
$G\equiv 
C_s(T_c)/\sqrt{\Delta U} = \-\sqrt{2/3}$ $(\pi^2$ $A^{-1}\-+1.5AB^2)$, 
$\xi^2=  \Delta
^{-2}(0)/2 \-= 3/ \-(16\Delta U)$,
$H\equiv (D-1)=$ $\Delta C/C_n(T_c)\-=1.5A^2\-B^2$ $/\pi^2$,
 and 
 ${\Delta U}\-/U_n(T_c)$  $= \-1.5 \-A^2/\pi^2$  
 where $\Delta C=C_s(T_c)-C_n(T_c)$ and 
 the universal constants $A$ and $B$ are defined by 
$A\equiv
\Delta(0)/T_c$ and 
$B^2= - [d\{\Delta(T)\-/\Delta(0)\}^2\-/d(T/T_c)]_{T=T_c}$. 
Unless the units of the variables are
explicitly mentioned,
 all energy (momentum)  variables are expressed in units of
$E_F$ ($k_F$),
 such that $\mu \equiv \mu/E_F$, $T \equiv T/T_F$, $q\equiv q/k_F$, $E_{\bf
q}\equiv E_{\bf q}/E_F$, $E_F=k_F=k_B=1,$ etc, where 
$\mu$ is the chemical potential and $E_F$ is the Fermi energy.
 The lengths  are expressed in units  of $k_F^{-1}$:
$\xi\equiv \xi k_F$.

In Sec. II we derive the present renormalized BCS and number equations.
 In Sec. III we present
an analytic study of the renormalized BCS equation and find several universal
relations among the observables. In Sec. IV we present a numerical study of
the present equation and establish power-law temperature dependence of some 
of the observables below $ T_c$ in non-$s$ waves.
 Finally, in Sec. V  we present a summary of our findings.

\section{Renormalized BCS and Number Equations\label{II}}

We consider a weakly
attractive short-range potential between electrons
in the angular momentum state $lm$,
\begin{equation}   
V_{\bf
p\bf q}=-V_0 g_{plm} g_{qlm} Y_{lm}(\Omega_p)Y_{lm}(\Omega_q),
\end{equation} 
where $g$ is the potential
form factor and 
$\Omega$ $(=\theta\phi)$represents
 the polar  and azimuthal  angles.  
  This potential leads to Cooper  instability for any
$V_0$ and $lm$. In even (odd) partial waves, pairing occurs in singlet
(triplet) state governed by the Cooper equation
\begin{equation}
V_0^{-1} = \sum_{{\bf q}(q>1)} g_{qlm}^2|Y_{lm}(\Omega_q)|^2
(2\epsilon_{q}-\hat E   )^{-1}, \label{120}\end{equation}
with  $B_c \equiv (2- \hat E)$ the Cooper binding, 
 $\epsilon_q =\hbar^2 q^2/2m$
 where $q$ is the wave number and $m$ the electron mass.

At a finite temperature, $T$, one has the  
following BCS gap and number equations for $N$ electrons
\begin{eqnarray}
\Delta_{\bf p}& =& -\sum_{\bf q} V_{\bf p \bf q}
\frac{\Delta_{\bf q}}{2E_{\bf q}}\tanh
\frac{E_{\bf q}}{2T} , \label{130}\\ N &=& \sum_{\bf q}
\biggr[1-\frac{\epsilon_q-\mu}{E_{\bf q}}
\tanh\frac{E_{\bf q}}{2T} \biggr], \label{140} \end{eqnarray} 
where 
$E_{\bf q} = [(\epsilon_q - \mu )^2 + |\Delta_{\bf q}|^{2}]^ {1/2},$ with
$\Delta_{\bf q}$ the gap function and $\mu$ the chemical potential. 
Though it is possible to have a BCS condensate in a mixed angular momentum
state, 
here  we assume, as in Ref. \cite{a},
 that the condensate is formed in a state of well-defined
$lm$, so that $\Delta _{\bf q}$ has the 
following anisotropic form: $\Delta _{\bf q} \equiv 
g_{qlm} \Delta_0 \sqrt{4\pi} Y_{lm}(\Omega_q)$ where 
   $\Delta_0$ and $g_{qlm}$  are dimensionless.  The  BCS gap is defined by
   $\Delta(T)=g_{q(=1)lm}\Delta_0$, which is the root-mean-square average of
   $\Delta _{\bf q}$ on the Fermi surface.
Equations
(\ref{120}) and (\ref{130}) lead to the renormalized BCS equation 
\begin{eqnarray}
\sum_{{\bf q}(q> 1)} \frac{|g_{qlm}Y_{lm}|^2}{2\epsilon_q- \hat E}-\sum_{\bf
q}
\frac{|g_{qlm}Y_{lm}|^2}{2E_q}\tanh \frac{E_q}{2T} =0.
\label{160} \end{eqnarray}
The summation is evaluated according to \begin{equation} \sum _{\bf q} \to
 \frac 
{N}{4\pi} \frac{3}{4}
\int_0^\infty \sqrt\epsilon_ q 
d\epsilon_ q \int d\Omega, \label{9}\end{equation}  where $\int d\Omega
=\int_0^{2\pi}d\phi
\int_0^\pi \sin\theta d\theta.$
Equations (\ref{140}) and (\ref{160}) can be explicitly written as
\begin{eqnarray}\int
d\Omega
\int d\epsilon_ q \sqrt \epsilon_q 
\biggr[1-\frac{\epsilon_q-\mu}{E_q}\tanh
\frac{E_q}{2T}\biggr] =\frac{16\pi}{3},\label{16}\end{eqnarray}
\begin{eqnarray}\int d\Omega |Y_{lm}|^2& \biggr[&
\int_{1}^\infty d\epsilon_ q  
\frac{\sqrt \epsilon_q g_{qlm}^2}{\epsilon_q-\hat E}- 
\int_0^\infty d\epsilon_ q 
\frac{\sqrt \epsilon_q g_{qlm}^2}
{E_q}\nonumber\\
&\times&
\tanh \frac{E_q}{2T}\biggr]=0. \label{15}
\end{eqnarray} 
The two terms in Eq. (\ref{16}) or (\ref{15}) under integral 
have ultraviolet divergences.
However, the difference between these two terms is finite. In the absence of
potential form factors ($g_{qlm}=1$), these
 equations are completely independent
of potential and are governed by the  observable 
 $B_c$. This is
why these equations are called  renormalized BCS equations \cite{c,re}. 
The quantity $B_c$   plays  the role of a 
potential-independent  coupling of interaction. 

Now we  calculate the critical temperature $T_c$  of
Eq. (\ref{15}), in the special case $g_{qlm}=1$. This potential is
independent of a range parameter and is usually called a zero-range
potential.
At $T=T_c$, $(\Delta(T_c)=0)$,  Eq. (\ref{15})  can be analytically
integrated to yield 
\begin{equation}
{T_c} = \frac{2{\exp(\gamma-1)}}{\pi}\sqrt{2 B_c}\approx 0.590 \sqrt{B_c}, 
 \label{21}
\end{equation} where $\gamma$ = 0.57722... is the Euler constant.
If $T_F$ is a 
few thousand  Kelvins, for a small $B_c$ in the weak-coupling region, one can
 have $T_c >$ 100 K appropriate for some high-$T_c$ materials. 
The standard BCS model yields in this case \cite{t}
\begin{equation}
\frac{T_c}{T_D}=\frac{\exp(\gamma)}{\pi}\sqrt{\frac{2B_c}{T_D}}. \label{22} 
\end{equation}

 To illustrate the advantage of Eq. (\ref{21}) over (\ref{22}) 
  in predicting a large $T_c$ in the weak-coupling limit, let us
consider a specific example  with $T_D=300$ K and $T_F=3000$ K.  
In the standard BCS
result (\ref{22}), the weak-coupling region is usually defined by 
$B_c\approx 1$ meV  or
$B_c/T_D\approx 0.037$. The smallness of $B_c$
justifies the weak-coupling limit and we take $B_c\le 1$ meV as defining the
weak-coupling region.   Schreiffer
\cite{8} suggested that $B_c$ is the proper measure of coupling. He noted
that $B_c=0.1$ meV is safely within the weak-coupling domain. 
 In this case for 
 $B_c = 1$ meV = 0.0037   
 one obtains from Eq. (\ref{22}) that $T_c$ is 
46 K. From Eq. 
 (\ref{21}), we obtain 
$T_c=0.036  =
107$ K. This result reflects an enhancement of $T_c$ in the renormalized model.
In order to provide further evidence of 
the weak coupling limit of the
present renormalized model with $T_c= 0.036$, we solved the number equation 
(\ref{16})  numerically
for the chemical potential $\mu$ and obtained $\mu=1.000$, which is
in the weak-coupling domain.  
The renormalized result (\ref{21}) has the advantage of  
producing  
the experimentally observed linear scaling between $T_c$ and $T_F$ in
high-$T_c$ materials \cite{u}.

\section{Analytic Study of the Renormalized BCS Equation}

There is no cut-off in the renormalized BCS equation (\ref{15}). At $T=0$ Eq.
(\ref{15}) can  be solved analytically in the absence of potential form
factors: 
$g_{qlm}=1$. 
Then 
each integral in Eq. (\ref{15}) is divergent at the
upper limit $\Lambda$, but for a sufficiently large $\Lambda$ the
difference becomes finite.  Now Eq. (\ref{15}) can be integrated in the
weak-coupling 
limit ($\mu = 1$) to yield: $$ 2\sqrt
\Lambda-\ln(e^2B_c/8)=2\sqrt \Lambda -2\ln[e^2\Delta(0) \sqrt{4\pi}/8]+\ln
F^2,
$$
where $\ln F=-\int d\Omega|Y_{lm}(\Omega)|^2\ln |Y_{lm}(\Omega)|$ with 
$e=2.718 281...$ 
For  $\Lambda
\to \infty$ this leads to   $ \Delta(0) =F \sqrt{2 B_c
}$ $/(e\sqrt \pi)$. However, $T_c$ is given by Eq. (\ref{21})  for all $lm$.
  In this case we have the universal constant 
$ A\equiv \Delta(0)/T_c =
F\sqrt \pi /\{2\exp(\gamma)\}.$

Though  $A$ is
independent of interaction model and dimension of space, 
 $\Delta(0)$ and $T_c$ are dependent on them. 
 For example, for a $s$-wave zero-range interaction 
 we have $\Delta(0)=\sqrt{2B_c}$ and
$T_c=\exp(\gamma)\sqrt{2B_c}/\pi$ from a renormalized BCS model 
in two dimensions\cite{c},  distinct from the above 
three-dimensional relations. For a fixed coupling, denoted by a  $B_c$,
we find an enhancement of $T_c$ in two dimensions over that in 
three dimensions by a factor of $e/2.$ In both two and three dimensions 
the renormalized BCS equation provides an enhanced $T_c$ over the standard
BCS $T_c$ given by Eq. (\ref{22}).

The entropy of the system is given by \cite{t}
\begin{equation}
S(T)=-2\sum_{\bf q}[(1-f_{\bf q})\ln(1-f_{\bf q})+
f_{\bf q}\ln f_{\bf q}],
\end{equation}
where $f_{\bf q}=1/(1+\exp( E_{\bf q}/ T))$.

The condensation energy per particle at $T=0$  is given by \cite{t}
$$
\Delta U\equiv {|U_s-U_n|}=\sum_{{\bf q}
(q<1)}2\zeta_q-\sum_{\bf q}(\zeta_{q}
-\frac{\zeta_q^2}{E_{\bf q}}-\frac{\Delta_{\bf q}^2}{2E_{\bf q}}), $$
where $\zeta_q=(\epsilon_q-\mu)$. 
In the absence of potential form factors
this can be evaluated  to lead to \cite{t}
$$
\Delta U = \frac{3}{8}\int d\Omega
\Delta^2(0) |Y_{lm}(\Omega)|^2,
$$
which yields $\Delta(0)/\sqrt{\Delta U}  = \sqrt{8/3}$ 
for all $lm$.  Using the
universal relation between $\Delta(0)$ and $T_c$,  one obtains 
  $T_c/
\sqrt{\Delta U}=\sqrt{8/3}A^{-1}$.
For all $lm$,  $U_n(T_c)= \pi ^2 T_c^2/4$, so that $\Delta U/U_n(T_c)
= 3A^2/(2\pi^2).$

The superconducting specific heat per particle  is given by
\begin{equation}
C_{s}= \frac{2}{NT^2}\sum_{\bf q}   f_{\bf q}(1-f_{\bf q})
\left( E_{\bf q}^2-\frac{1}{2}T\frac{d\Delta_{\bf q}^2}{dT} \right).\label{sp} 
\end{equation} 
The normal specific heat $C_n$ is given by Eq. (\ref{sp}) with $\Delta_{\bf q}
=0.$
The jump in specific heat  per particle 
at $T=T_c$ ($\Delta(T_c)=0$), $\Delta C \equiv  [C_s-C_n]_{T_c}$
 is given by
\cite{t}
\begin{eqnarray}
\Delta C =-\frac{1}{NT_c} \sum_{\bf q}\left[f_{\bf q}(1-f_{\bf q})
\frac{d\Delta_{\bf q}^2}{dT}
\right]_{T_c}. 
\label{dc}\end{eqnarray}
In the special case $g_{qlm}=1$, the radial integral in Eq. (\ref{dc})
can be evaluated as in Ref. \cite{t} and we get 
\cite{t}\begin{eqnarray}
\Delta C& = & -\frac{3}{4T_c} \int  d\Omega
\int {\sqrt \epsilon_q
d\epsilon_q }\nonumber \\ &\times &\biggr[f_{\bf q}(1-f_{\bf q})
 \frac{ d\Delta^2(T)}{dT}\biggr]_{T_c}|Y_{lm}(\Omega)|^2 .   \end{eqnarray}
This
 leads to \cite{t}
  $\Delta C= -(3/4)[d\Delta^2(T)/dT]_{T=T_c}= (3/4)A^2 B^2 T_c$ 
  for  all $lm$.
From this,  we obtain
$H\equiv (D-1)=\Delta C/C_n(T_c)\equiv 1.5A^2B^2/\pi^2,$  where 
 $C_n(T_c)=\pi^2T_c/2 $, so that
$C_s(T_c)= (\pi^2+1.5A^2B^2)T_c/2$. Consequently, 
$C_s(T_c)/\sqrt{\Delta U}
\equiv G = \sqrt{2/3}(\pi^2A^{-1}+1.5AB^2)$. 
The numerical values of the constants $A$, $B$, $H$, $F$ and $G$ are
given in Table I.

The spin-susceptibility $\chi$ of the system is defined by 
\cite{a}
\begin{equation}
\chi(T)= \frac{2\mu_N^2}{T}\sum_{\bf q}f_{\bf q}(1-f_{\bf q}),
\end{equation}
where $\mu_N$ is the  nuclear magneton. At $T=T_c$, $\chi_s(T)
=\chi_n(T)$ and it is appropriate to study the ratio $\chi_s(T)/\chi_n(T_c)$.

Finally, it is also of interest to study the penetration depth $\lambda$
defined by \cite{t}
\begin{equation}
\lambda^{-2}(T)=\lambda^{-2}(0)\left[ 1-\frac{2}{NT} 
\sum_{\bf q}f_{\bf q}(1-f_{\bf q})\right].
\end{equation}
In the numerical study  of next section we shall calculate $\Delta\lambda
(T)= (\lambda(T)-\lambda(0))/\lambda(0).$

The dimensionless $s$-wave pair radius
 defined by $\xi ^2= \langle \psi_q|r^2 |\psi_q\rangle/
\langle \psi_q|\psi_q\rangle$, with the pair wave function $\psi_q= 
g_{qlm}\Delta
/(2E_q)$, can be evaluated  by using $r^2 = -\nabla_q ^2$.
In the weak-coupling limit,
the zero-range analytic result of Ref. \cite{c} leads to $\xi^2=  \Delta
^{-2}(0)/2 = 3/ (16\Delta U)$. Consequently,
\begin{equation}\label{tc}
\xi= \frac{1}{\sqrt 2 A T_c}
\end{equation}

\section{Numerical Study}

Equations (\ref{16}) and (\ref{15}) are solved numerically 
without  approximation in $s$, $p$ and $d$
waves for separable potentials with dimensionless
form factors $g_{qlm} =\epsilon_q^{l/2} [\alpha/(\epsilon_q
+\alpha)]^{(l+2)/2} $ 
with correct threshold behavior as $q\to 0$, where
$\alpha $ is the range parameter. (Normally, one uses in Eq. 
(\ref{9})
$d\epsilon_q \sqrt \epsilon_q = d\epsilon_q \cite{8,t}.)$ 
Following Refs. \cite{t,a},
we calculated  
 $\Delta (0)$, $T_c$, $C_s(T_c)$, 
the $s$-wave pair radius $\xi^2$ at $T=0$ as well as $\Delta(T)$,
$\lambda(T)$, $C(T), S(T)$, and $U(T)$ 
for different $V_0$ and $\alpha$. In Fig. 1 we plot 
$\Delta (0)$, $T_c$, $C_s(T_c)$, and 
 $\xi^2$ versus $\Delta U$ and establish  universal scalings mentioned
before.  The calculations were repeated 
by  varying $\alpha  $ from 1 to $\infty$ and we found  Fig. 1  to be
insensitive to this variation for each $lm$. For $l\ne 0,$ Eqs.  (\ref{16})
and (\ref{15}) diverge for $\alpha \to \infty$ and calculations were
performed for $\alpha = 1 $ to 10. The increase in $\Delta U$ of Fig. 1
corresponds to an increase in coupling $V_0$.  We could plot the variables of
Fig. 1 in terms of $V_0$ as in Ref. \cite{No}. Then each $\alpha$
 leads to a distinct curve. However, if we express the variation in
$V_0$ by a variation of an observable of the superconductor, such as
$\Delta U$ or $T_c$, universal potential-independent scalings are obtained.
In each case the exponent and prefactor of each scaling relation are in
excellent agreement with the analytic result obtained above without form
factors.

\vskip .2cm

{TABLE I. Numerical values of various constants and exponents
in different angular momentum
states}

\vskip .2cm

\begin{centering} 

\begin{tabular} {|c|c|c|c|c|c|c|c|c|c|} \hline  &  &
  &  &   & & $\beta_S$ &$\beta_{C}$ &
  $\beta_\chi$& $\beta_\lambda$\\
 $lm$&$F$ &$A$ &$B$ &$H$ &$G$ &$\approx$&$\approx$&$\approx$&$\approx$\\
 \hline 
 00 &3.5449 & 1.764 &1.74& 1.43 &11.11 &  && &\\ 
 10 & 2.8563 & 1.422 & 1.60& 0.79&10.12&  2&   2&& 1.1\\ 
 11 & 3.3300 & 1.658 & 1.68& 1.18 &10.59&3&2.6& &2.4\\ 
 20 & 2.7748 &1.382 & 1.57& 0.72 &10.00& 2& 2&1.2 & 1.1\\ 
 21 & 3.1006 &1.544 & 1.63& 0.96 & 10.24&2.1& 2& 1.4& 1.5\\ 
 22 & 3.1006 &1.544 & 1.63& 0.96 & 10.24& 2.1 & 2& 1.4& 1.5\\ 
\hline 
\end{tabular}

\end{centering}

The $T_c$ should not arbitrarily increase with coupling as Fig. 1 may imply.
With increased coupling the electron pairs form composite bosons which 
undergo Bose condensation below $T=T_c
\equiv 0.218 $, for bosons with twice the electron mass and half the
electron density
\cite{No}.  Hence the $T_c$ curve of Fig. 1 is only plotted up to 
about 
  $T_c = 0.1$. For a large class of unconventional three-dimensional 
 superconductors $T_c$ has been 
  estimated to be  0.05 \cite{u}, 
where the universality of the present study should hold. For a typical 
high-$T_c$ material $T_c = 0.04$ and from Eq. (\ref{tc}) 
we find pair-size $\xi\approx 10$  in $s$ wave. Hence with the increase of
$T_c$, $\xi$  has reduced appropriately in the weak coupling region as 
found experimentally.

Next we studied the temperature dependence of $\Delta(T)$, $S_s(T)$,
$C_s(T)$, $\chi_s(T)$, $U_s(T)$, and $\lambda_s(T)$ for $T< T_c$ for
different  $V_0$, and range $\alpha $ varying from 1 to $\infty$. We found
that  $\Delta(T)/\Delta(0)$ versus $T/T_c$ is an universal function for each
$lm$ independent of potential parameters. We find the  universal fit
$\Delta(T)/\Delta(0)\approx B (1-T/T_c)^{1/2}$ valid for $T\approx T_c$ with
numerical values of $B$ quoted in Table I.  For $s$-wave BCS superconductors
$S_s(T)$, $C_s(T)$, $\chi_s(T)$, and $\lambda_s(T)$ have exponential
dependence on $T$ as $T \to 0$. But for non-$s$ wave states, these variables
have power-law dependence on $T$ as observed in some materials
\cite{t2,s,t1}. In Figs. 2, 3, 4, and 5 we plot $S_s(T)/S_n(T_c)$,
$C_s(T)/C_n(T_c)$, $\chi_s(T)/\chi_n(T_c)$, and $\Delta
\lambda(T)/\lambda(0),$ respectively, versus $T/T_c$  where $\Delta \lambda
(T)=(\lambda(T)-\lambda(0))/\lambda(0).$ As commented in Ref. \cite{a},
$\chi_s$ will be significantly different from $\chi_n$ only for even $l$.
 We have calculated
$\chi_s$ only for $l=0$ and 2.  Scalings (\ref{a}) $-$ (\ref{d}) are
established in Figs. 2 $-$ 5.  The  exponents of these scalings are given in
Table I. In order to find $\beta_C$ we also plotted $C_s(T)/C_n(T_c)$ versus
$T/T_c$ on log scale. That plot was used to calculate the exponent $\beta_C$.
However, on log scale different curves nearly overlap and
hence that plot is not shown here. From Fig. 3 we find that
the zero of $[C_s(T)-C_n(T)]$ appears 
in all cases for $T/T_c \approx 0.5.$ Moreover, all curves for 
 superconducting 
specific heat  meet at $T/T_c \approx 0.6$. These two behaviors seems 
to be  typical for  models based on BCS equations.

The  constants in Table I  for different $lm$ satisfy ${\cal C}_{00}>{\cal
C}_{11}>{\cal C}_{21}={\cal C}_{22}>{\cal C}_{10} >{\cal C}_{20}$, where
${\cal C}_{lm}$ stands for  $F$, $A$, $B$, $H$, and $G$.  Hence the following
sequence of $lm$ states represents the increase of anisotropy for the gap
function: 00, 11, (21,22), 10, and 20. From the plot of entropy in Fig.  2,
we find that this sequence of $lm$ also represents the increase of disorder
and consequently,  a decrease in superconductivity or an approximation to the
normal state, as is clear from Figs.  3 $-$ 5.  Because of approximation to
more anisotropy and disorder, the observables for the normal state are closer
to the superconducting $l\ne 0$ state  than to the superconducting $l=0$
state.

The exponents $\beta_S$, $\beta_C$, $\beta_\chi$ and $\beta_\lambda$
are critical exponents  near $T=T_c$.  Wilson \cite{w}
discussed the universal nature of similar critical 
exponents in ferromagnetism and
concluded that the numerical value of those exponents depend on the
dimensionality of space and the symmetry of the order parameter of phase
transition.    Recently, we have calculated 
some of these exponents using the renormalized BCS equation in 
 two dimensions \cite{two}.
From these studies it seems  that these universal critical exponents of BCS
superconductivity are also determined by the
dimensionality of space and the symmetry of the order parameter  $\Delta_{\bf
q}$.

\section{Conclusion}
Through a numerical study of the renormalized weak-coupling
BCS equation in three dimensions in $s$, $p$ and $d$ waves we have
established robust scaling of   $\Delta(0)$, $T_c$, $C_s(T_c)$, and $\xi^2 $,
as a function of   $\Delta U$, independent of range of a general separable
potential.  The $T$ dependence of $S_s(T)$, $C_s(T)$, $\chi_s(T)$, and
$\Delta\lambda(T)$  below $T_c$ in non-$s$ waves show  power-law scalings
distinct to some high-$T_c$ materials \cite{t2,s,t1}. No power-law $T$
dependence is found in $s$ wave for these observables.  The universal nature
of the solution does not essentially change with the potential range and 
remains valid for a zero-range potential.  
In the
weak-coupling limit  the present solutions of the renormalized BCS equations
simulates typical high-$T_c$ values for the coherence length $\xi$, and
$T_c$.  They also exhibit the $T_c$ versus $T_F$ linear correlation (at a
fixed $B_c$)  as observed by Uemura \cite{u}.

We thank John Simon Guggenheim Memorial Foundation, Conselho Nacional de
Desenvolvimento Cient\'{\i}fico e Tecnol\'ogico and Funda\c c\~ao de Amparo
\`a Pesquisa do Estado de S\~ao Paulo for financial support.

Figure Captions:
\vskip .2 cm

1. $C_s(T_c)$ (dashed line),  $T_c$ (dotted line), $\Delta(0)$ (dashed-dotted
line) for different $lm$ and $s$-wave pair radius $\xi^2$ (solid line)
versus zero-temperature condensation energy $\Delta U$ for different $V_0$
and  $\alpha$ (from 1 to $\infty$). For $C_s(T_c)$ and   $T_c$ there are six
distinct lines and for $\Delta(0)$  we have a single line for all $\alpha$
and $lm$. The lines  for $C_s(T_c)$ ($T_c$)  correspond to $lm= 00, 11,
(21,22), 10,$ and 20 from top to bottom (bottom to top).

2. Entropy $S_s(T)/S_s(T_c)$  versus $T/T_c$ for different $lm$, $V_0$, and
$\alpha $ between 1 to $\infty$. The curves are labelled by $lm$.

3. Same as Fig. 2 for specific heat $C_s(T)/C_n(T_c)$  versus $T/T_c$.

4. Same as Fig. 2 for spin-susceptibility $\chi_s(T)/\chi_s(T_c)$ versus
$T/T_c$.

5. Same as Fig. 2 for $\Delta\lambda(T)$  versus $T/T_c$.

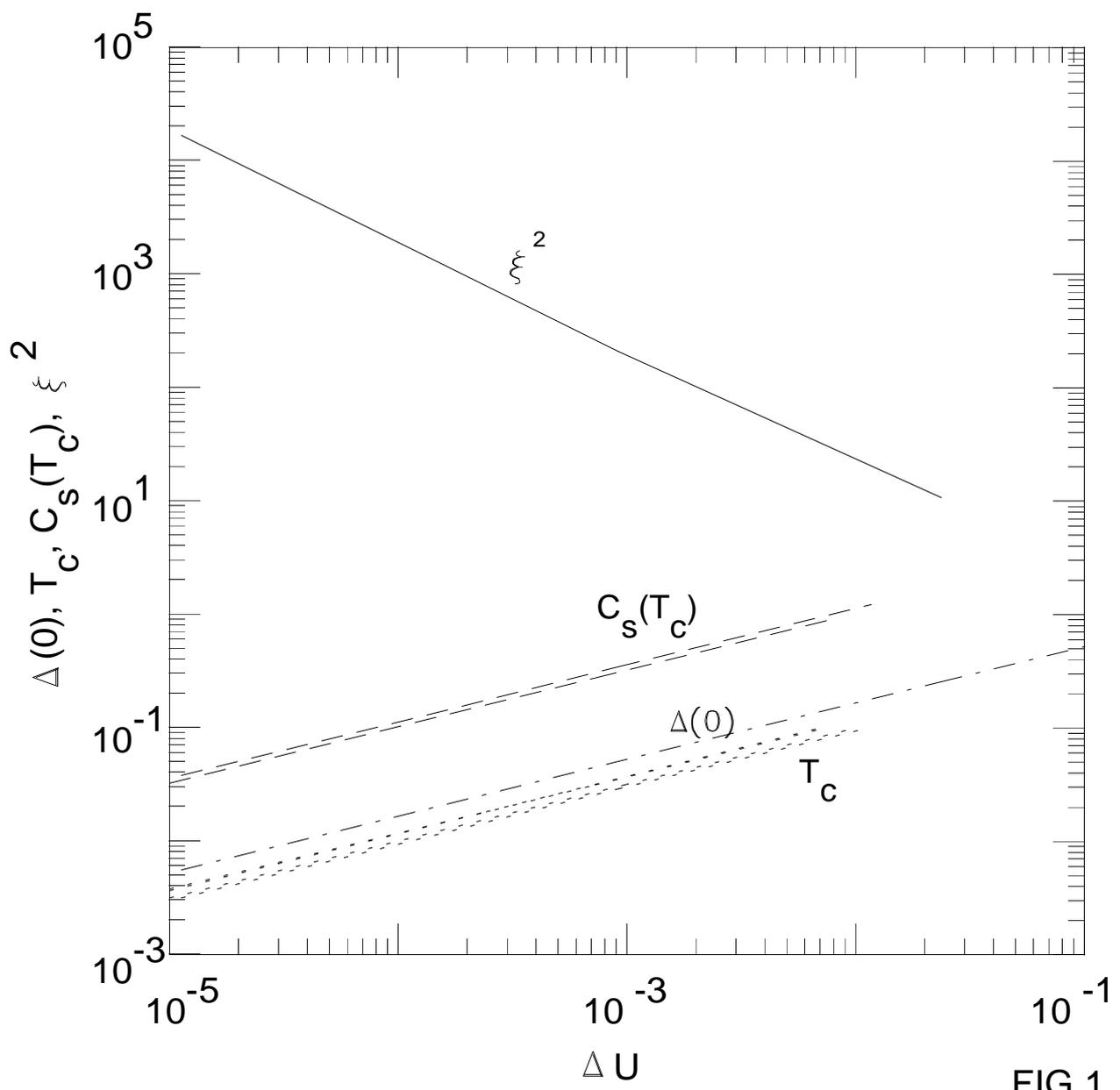
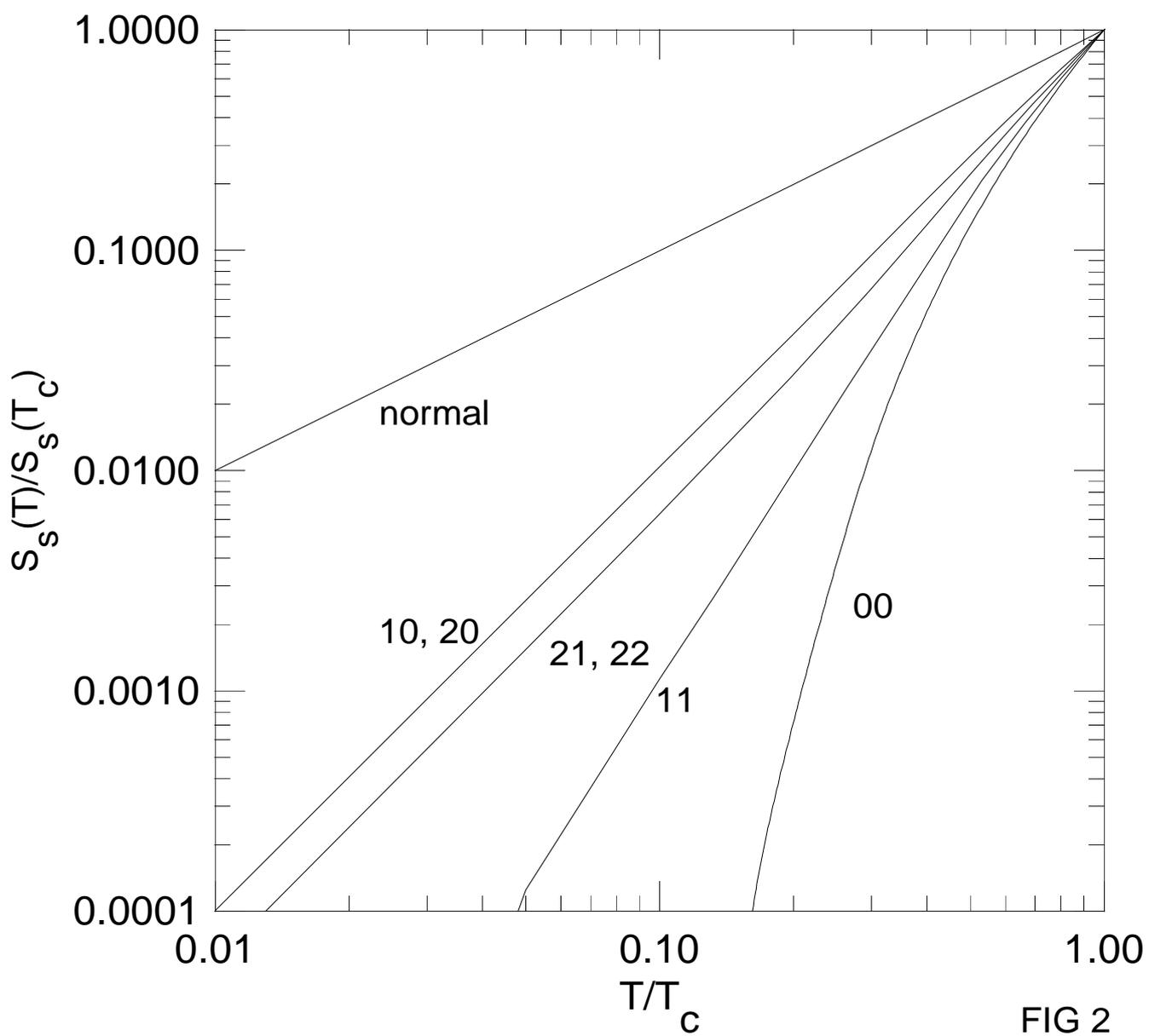
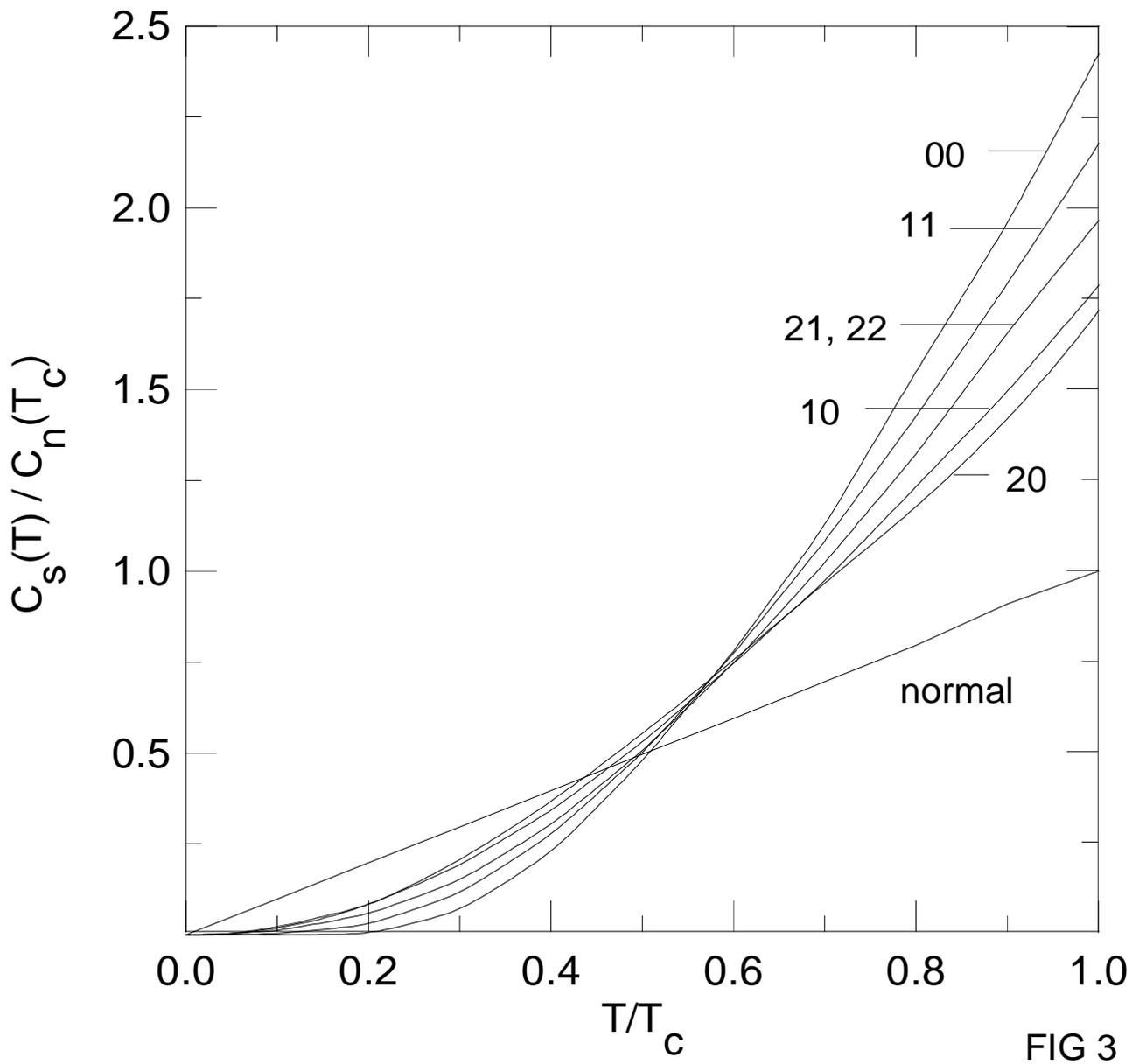
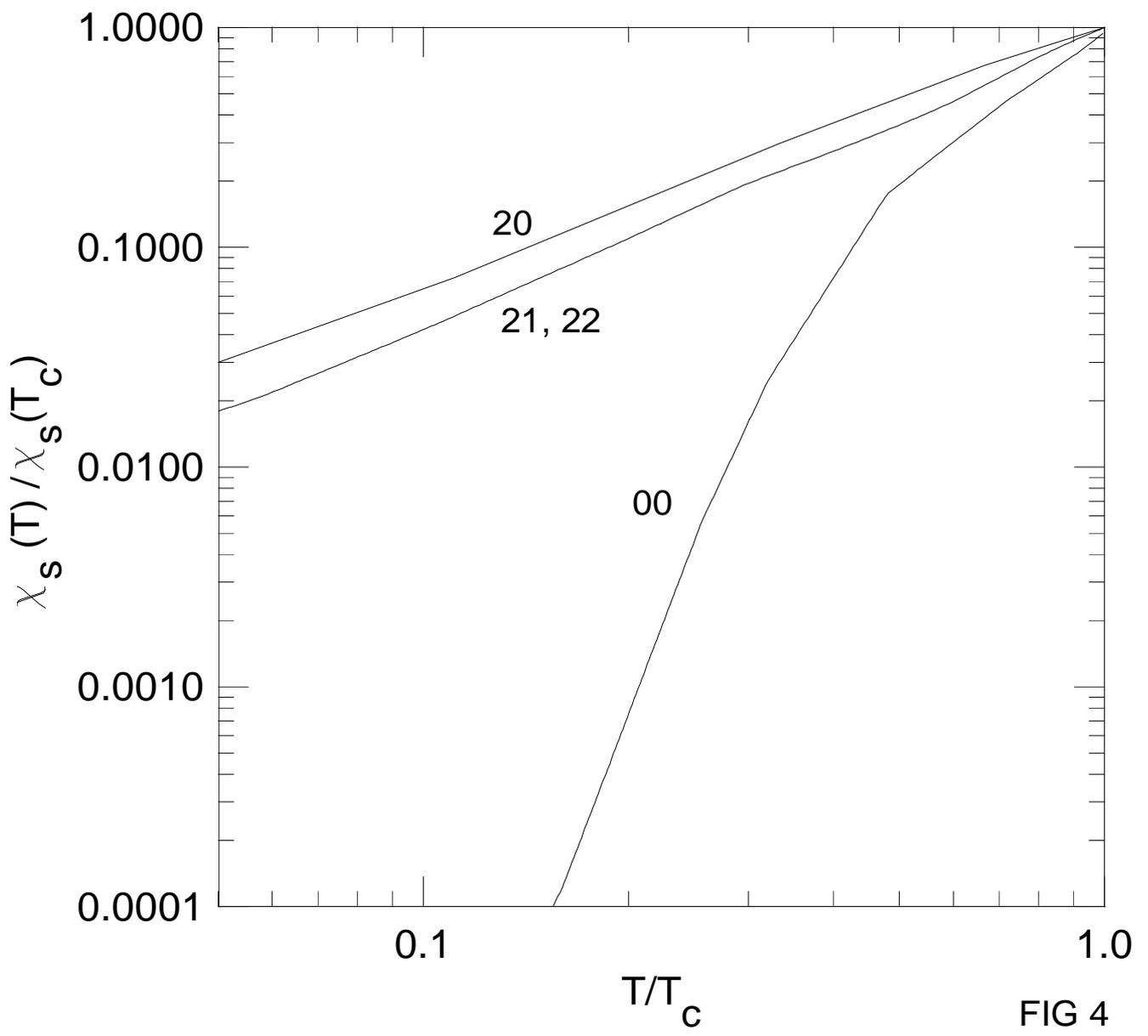
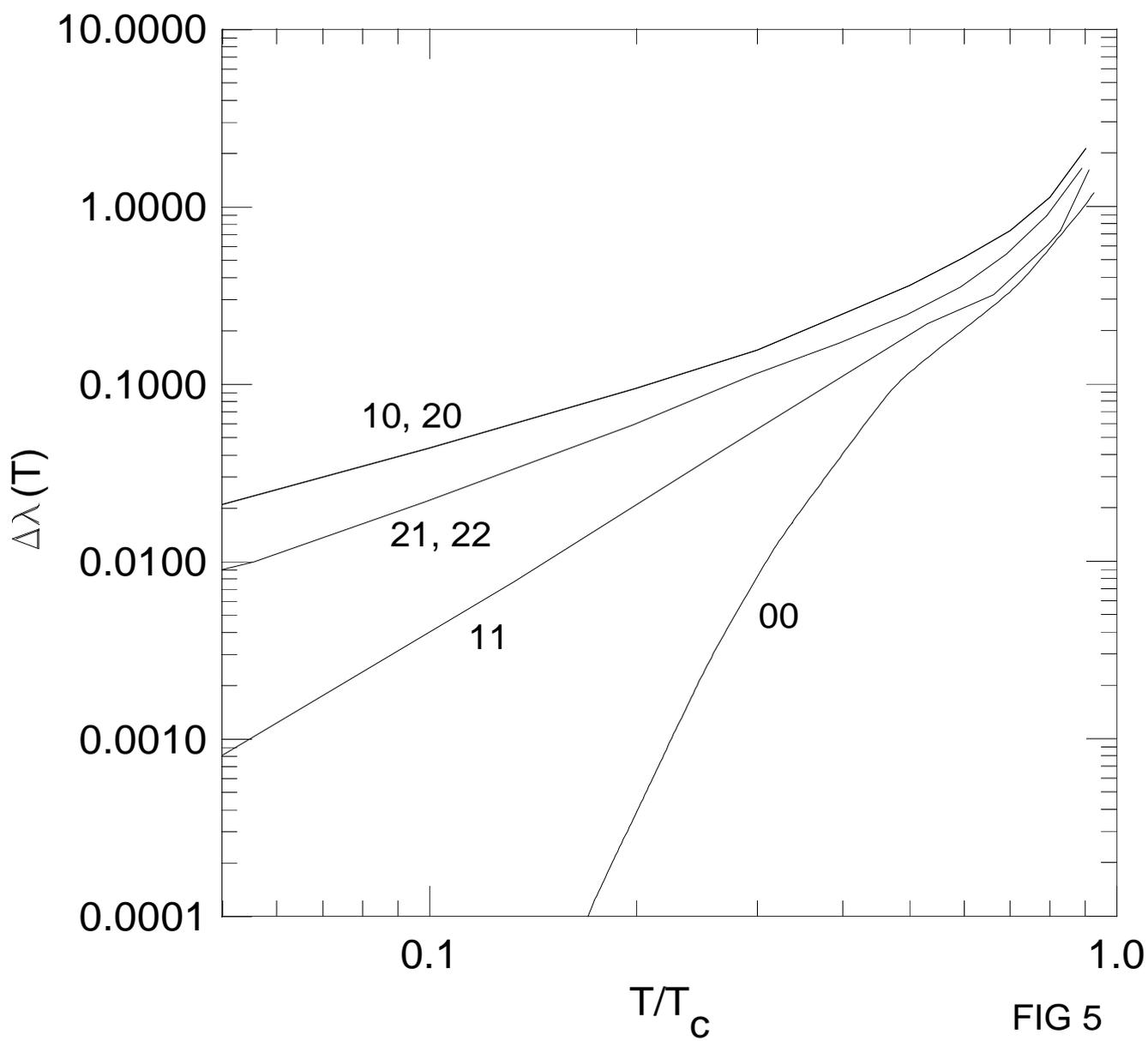
\end{document}